\documentclass[12pt,preprint]{aastex} 
\usepackage{emulateapj5}

\newcommand{\be}{\begin{equation}}
\newcommand{\ee}{\end{equation}}
\newcommand{\ba}{\begin{eqnarray}}
\newcommand{\ea}{\end{eqnarray}}

\newcommand{\sgrid}{{s_{\rm grid}}}
\newcommand{\bld}[1]{\mbox{\boldmath$#1$\unboldmath}}
\newcommand{\edot}{\dot{\cal{E}}}
\begin{document}
\shortauthors{Lithwick \& Chiang}  
\shorttitle{Collisional Disks}

\title{Collisional Particle Disks} 

\author{Yoram Lithwick\altaffilmark{1}
\& Eugene Chiang\altaffilmark{2,3}}
\altaffiltext{1}{Canadian Institute for Theoretical Astrophysics,
60 St.~George Street,
Toronto ON M5S 3H8, Canada}
\altaffiltext{2}{Astronomy Department,
University of California at Berkeley,
Berkeley CA~94720, USA}
\altaffiltext{3}{Alfred P.~Sloan Research Fellow}  
  
\email{yoram@cita.utoronto.ca, echiang@astron.berkeley.edu}  

\begin{abstract}
We present a new, simple, fast algorithm to numerically evolve
disks of inelastically colliding particles surrounding a central star.
 Our algorithm adds negligible computational cost
to the fastest existing collisionless N-body codes, and can be used
to simulate, for the first time, the interaction of planets with disks
over many viscous times.
Though the algorithm is implemented in two dimensions---i.e., the motions
of bodies need only be tracked in a plane---it captures the behavior of
fully three-dimensional disks in which collisions  maintain inclinations that are comparable
to random eccentricities.
 We subject the algorithm to a battery of tests
for the case of an isolated, narrow, circular ring. Numerical simulations
agree with analytic theory with regards to how particles' random velocities
equilibrate; how the ring viscously spreads; and how energy dissipation,
angular momentum transport, and material transport are connected.
We derive and measure the critical value of the coefficient of restitution
above which viscous stirring dominates inelastic damping and the particles'
velocity dispersion runs away.

\end{abstract}

\section{Introduction}

How does a disk of collisional particles surrounding a star evolve in the presence of planets?
The answer to this  question has important implications.
  For example, after
the planets of our Solar System accreted most of their mass,  many small, rocky and icy bodies remained orbiting the Sun.
Somehow, the planets eliminated most of these remnant planetesimals, while leaving some behind to form the asteroid belt, the Kuiper belt, and the Oort cloud.  
In the vicinity of Uranus and Neptune, the small bodies must have been
highly collisional. Otherwise,
these planets would have taken $10^{12}$ yr to form {\it in situ} \citep{GLS04}.\footnote{
In the terrestrial zone, the small bodies were also likely collisional, although the case is not
as convincing there as it is in the outer Solar System \citep{GLS04}.}  Yet virtually all simulations of the
late stages of planet formation in the outer Solar System---such as those that model the
migration of the ice giants, the resulting trapping of Kuiper belt objects into resonances,
and the ejection of small bodies to the Oort cloud---neglect
 collisions.
When the effects of collisions are accounted for, the current picture of the formation of planetary systems might change drastically.

Planetary rings provide another setting in which
interparticle collisions play a crucial role.
What are the origins of narrow rings
shepherded by satellites? How do narrow rings settle into their
special apse and node-aligned states \citep[e.g.,][]{CC04}?
And how do rings back-react upon and shape the orbits of shepherd
satellites? Our understanding
of satellite-ring interactions bears on mysteries such as the origin
of eccentricities of extra-solar giant planets
\citep[e.g.,][]{GS03}.

Despite its importance, the behavior of particle disks in the presence of
 perturbing bodies is poorly understood.  
Numerical simulations can help to further understanding. 
But until now, simulations of collisional disks have been too inefficient to follow, say, how disks viscously spread in the long term.
 Collisions are traditionally simulated with a brute-force method
\citep[e.g.,][]{Bra77,WT88}: at each time step of the integration of 
the gravitational equations of motion, it is determined
which pairs of particles might collide before the next timestep.  These potential collision pairs 
   are 
then integrated forwards in time with a much smaller timestep, to see if they really do collide.   
But this method is inefficient:
a brute-force search for collision partners requires around $N_{\rm tp}^2$ operations at each timestep, where $N_{\rm tp}$ is the number of test particles.  In addition, most potentially colliding pairs do not collide, particularly in optically thin disks.  Hence much computing time is wasted on missed collisions. 
More complex algorithms have been devised to reduce computing time \citep[e.g.,][]{LS00,CTB01}.  But
these are still not nearly as fast as the fastest collisionless N-body codes, such as
SWIFT \citep{LD94}.  

We sought a collision algorithm that (i) could be added to any N-body code, such as the
freely-available SWIFT; (ii)
contributes negligibly to the computational cost; (iii) is simple conceptually; 
(iv) is easy to code; and (v)  
follows correctly the long-term viscous evolution of disks in the presence of planets.
 We designed our algorithm to simulate
 a vertically optically thin disk of identical, collisional, massless, inelastic but indestructible test particles that feel the gravity of the Sun and of multiple planets.
  Complications that we do not include, such as the self-gravity of the particles,
 order-unity optical depths, and particles with differing sizes, spins, and cohesive strengths, could all affect the viscous
 evolution in ways that are not currently understood. But at this stage
 it seems wisest to
 ignore these complications, even though the algorithm could be modified to handle
 them. Viewed in its most basic terms, inelastic collisions
dampen random velocities and act as a source of friction between neighboring streamlines.
As long as our algorithm preserves this behavior, while conserving angular momentum and accounting
for the loss of energy in inelastic collisions, it seems likely that it will properly model the long-term evolution of
collisional disks.  In the present paper, we test this assertion thoroughly when there are no planets, comparing in detail the results of our simulations with those of analytic theory. In a future paper,  we shall test our algorithm in
the presence of planets.
 
\section{The Collision Algorithm}

The gravitational equations of motion of
the Sun, planets, and  massless 
test particles are integrated
with the Wisdom-Holman mapping method \citep{WH91},
using the SWIFT subroutine package \citep{LD94}.  

We supplement SWIFT with a subroutine that simulates collisions between test particles in 
a disk with vertical optical depth
\be
\tau\sim N_{\rm tp}{s^2\over \bar{r}\cdot \Delta} < 1 \ , \label{eq:taunew}
\ee
where  $N_{\rm tp}$ is the number of test particles, $s$ is their size, and
$\bar{r}$ and $\Delta$ are, respectively,
the mean orbital radius and the radial width of the annulus that the particles occupy.
In collisional particle disks, collisions tend to isotropize the velocity distribution.\footnote{More
precisely, in optically thin disks the r.m.s. azimuthal speed is twice the r.m.s. radial speed;
the r.m.s. vertical speed is comparable.}
  The collision time is $t_{\rm col}\sim 1/(n_{\rm v}s^2u)$, where
$u$ is the 1-D random speed and $n_{\rm v}$ is the volumetric number density, which is related to $\tau$
via $\tau\sim n_{\rm v} s^2 u t_{\rm orb}$.
Therefore
\be
t_{\rm col} \sim \frac{t_{\rm orb}}{\tau} \, . \label{eq:tcolnew}
\ee
  The collision time is longer than the orbital time by the $u$-independent
factor $1/\tau$.

We capture this behavior with two-dimensional simulations in which
all bodies have zero inclination.
Every time step $dt$, a two-dimensional square grid is built, with each grid element having
dimensions $\sgrid\times\sgrid$; $\sgrid$ can be thought of as the size of a particle.  If two test particles fall in the same grid cell, 
and if their relative speed is negative (i.e., if they are
approaching each other), then they collide with each other with probability
 $P_{\rm col}= dt/t_{\rm orb}\ll 1$, where $t_{\rm orb}$ is the orbital time at the collision point.
A random number generator is used to determine whether or not they
actually collide. 

To see that this algorithm gives the same collision time as Equation (\ref{eq:tcolnew})
(where $\tau$ is given by Eq. [\ref{eq:taunew}] with $s\rightarrow s_{\rm grid}$), it is instructive to
consider first a simpler algorithm that also yields the correct collision time.
  In this simpler algorithm, one waits for
a time interval of $t_{\rm orb}$ (instead
of $dt$)
before finding which particles fall in the same grid cell. Then two particles which do fall 
in the same grid cell, and have converging velocities, collide with probability $P_{\rm col}= 1$. 
Since the probability that a given particle lies in a cell occupied by a second particle is
$\tau$, the collision time is $t_{\rm orb}/\tau$, as required.\footnote{Although this simpler algorithm yields the correct collision time, we did not use it because it introduces
an artificial frequency into the problem, set by the time interval at which the algorithm is applied
($\sim t_{\rm orb}$). When we attempted this algorithm, we found that a gap was cleared in the disk
of test particles where the orbital period was exactly equal to this interval.}
Turning now to the algorithm that we actually use,  since we apply this algorithm every time interval $dt$
(and not $t_{\rm orb}$),
we must correspondingly reduce the probability of a collision by $dt/t_{\rm orb}$ in order to maintain 
the collision time at the value given by Equation (\ref{eq:tcolnew}).

Although carried out in only two dimensions, we emphasize that our algorithm 
models three-dimensional disks in which collisions maintain 
inclinations that are comparable to the random eccentricities.
A truly two-dimensional disk is not realistic because collisions invariably generate out-of-plane velocities.
But if one could somehow prevent the generation of out-of-plane velocities, the collision time in such
a disk would be 
 $\sim s/(u\tau)$, which differs
from Equation (\ref{eq:tcolnew}) by the factor $s/(u t_{\rm orb})$. Since our algorithm satisfies Equation 
(\ref{eq:tcolnew}), it does not model truly two-dimensional disks.

As will be shown below, collisions 
drive the
random speed of the particles to
$u\gtrsim  \sgrid/t_{\rm orb}$.  Hence if two particles fall in the same grid cell at one time, they will usually fall in separate grid cells
after one orbital period.   Since collisions can potentially occur every time step, it might be thought that the same two particles can collide many times in succession---a behavior that we consider undesirable.  But this behavior is avoided by the requirement that particles must be approaching each other for a collision to occur; immediately after they collide, their relative velocity reverses sign, and they are no longer candidates for a collision pair.

One of the main advantages of our algorithm is that the timestep is not restricted by the Courant
condition.  In a brute-force algorithm, one must restrict $dt\ll s/u$ in order to ensure that any two particles
that fall within a distance $s$ of each other collide.  This restriction on $dt$ can be very cumbersome when
$u\gg s/t_{\rm orb}$, as it will be whenever planets stir up the eccentricities.  We avoid the Courant condition
by treating the vertical dimension statistically: when two particles fall within the same two-dimensional grid cell, they need only collide a small fraction of the time because their vertical positions will, in general, differ.
With our algorithm, we may choose $dt$ to be as large as is allowed by SWIFT, which is typically a significant fraction of the orbital time.

If two particles have been selected for a collision, i.e., if they lie in the same grid cell, are approaching each other, and are selected by the random number generator, then
their velocities are updated as though the bodies were
frictionless spheres whose surfaces touch \citep[e.g.,][]{Tru71}:
the component of the relative velocity vector that
lies parallel to the axis connecting the two particles is reversed in sign (from a converging 
velocity to a diverging one), and
multiplied by the coefficient of restitution $\epsilon$, i.e., in obvious notation, 
\be
u_{\rm rel,\parallel}'=-\epsilon u_{\rm rel,\parallel}  \ .
\label{eq:upar}
\ee  
Neither 
the perpendicular component
of the relative velocity vector, nor
the velocity of the center of mass of the two colliders, nor the positions of the colliders are changed by the collision.
A collision does not alter the sum of the angular momenta of both colliding bodies; hence the
collision algorithm exactly conserves total angular momentum. 
Note that a collision between two particles separated by distance $d$ changes the velocities
of the particles as though each was a 
 smooth sphere with radius $d/2$.  
Since $d$ changes from collision to collision, the particles' sizes are effectively changing; they 
are 
only approximately $s_{\rm grid}$.

The algorithm has now been completely described, aside from how 
the code 
finds which 
pairs of particles lie in the same grid cell.   To find colliding pairs,  the code first determines
in which grid cell each particle lies.  A grid cell is labelled by two integers, representing
its location along the x- and y-axes.   Second, the code sorts the grid cells that contain test particles with the heapsort algorithm \citep{PTV+92}. The sorted occupied grid cells are then checked to see if the same grid cell is repeated for two different particles.  The step that takes the most time in the entire collision algorithm is the heapsort, which requires 
$\sim N_{\rm tp}\ln N_{\rm tp}$ operations.  However,  in the runs presented in this paper, 
with  $N_{\rm tp}=10^4$ particles,
it was found that
the collision algorithm contributed negligibly to the running time of SWIFT.

\section{Simulations of Narrow Circular Rings}

In the remainder of this paper, we investigate  circular rings of particles, without
any planets.  
Circular rings are understood quite well theoretically \citep[e.g.,][]{LP74,Bra77,GT78,Shukhman84,PH87}.
Our goal is not only to test the collision algorithm, but also to develop
diagnostics that can be used for the much more complicated 
case when planets are present.

The parameters of the simulation include the coefficient of restitution ($\epsilon$), 
the size of a grid element ($\sgrid$), the number of test particles ($N_{\rm tp}$), and the initial orbital elements of the test particles.
The central body's mass is $1 M_\odot$.  We simulate narrow rings with mean radius
$\bar{r}=$ 1 AU and radial width $\Delta \ll \bar{r}$, and choose $dt= 0.18$ years for the integration timestep.

A particle ring has two characteristic timescales \citep{Bra77}.
 The shorter one is $t_{\rm col}$, the time for 
a particle to collide.
The longer one is  $t_{\rm diff}$, the time for particles to diffuse across the 
ring's width.
Random velocities relax to their equilibrium distribution on timescale $t_{\rm col}$, while
the ring density evolves on timescale $t_{\rm diff}$.
We investigate in turn the evolution of random velocities and of density.

\section{Random Velocity Evolution}
\subsection{Theory}

Collisions can both excite random velocities by drawing energy from
the background Keplerian shear (``viscous stirring''), and damp
random velocities because of finite inelasticity.
When the coefficient of restitution $\epsilon < \epsilon_*$, 
where $\epsilon_*$ is a critical value, a typical collision
damps random velocities. But there is a limit to how cold
the particles get.
Particles on circular orbits collide with one another at the Keplerian
shearing speed $\sim s\Omega$, where
$s$ is the size of a particle and $\Omega$ is the orbital angular frequency.
In our collision routine, this relative speed is $\sim s_{\rm grid}\Omega$.
Since a single such collision redirects the particles
onto non-circular orbits,
the random velocity cannot fall below $\sim s_{\rm grid}\Omega$, and the rms (root-mean-squared) eccentricity always relaxes to
\be
e_{\rm rms}\sim {s_{\rm grid}\over r}
\label{eq:emin}
\ee
in the $\epsilon \ll \epsilon_*$ limit,
where $r$ is the local disk radius.

By contrast, when $\epsilon > \epsilon_*$, a typical collision excites
random velocities: viscous stirring dominates inelastic damping,
and the rms eccentricity runs away.

In optically thin disks composed of equal-size particles, $\epsilon_*$ is determined solely by  the angular dependence of the 
differential collisional cross-section.  For frictionless spheres, $\epsilon_*=0.63$ \citep{GT78}. For frictional spinning spheres, $\epsilon_*=0.92$ \citep{Shukhman84}.   For our collisional cross-section,  we derive
$\epsilon_*=\sqrt{7}/5=0.529$ (Eq. [\ref{eq:star}]).

\subsection{Simulations: Approach to Velocity Equilibrium}
\label{sec:approach}
Figure \ref{fig:e_vs_t} shows the time evolution of the rms eccentricity,
\be
e_{\rm rms}\equiv \langle e^2\rangle^{1/2} \,,
\ee
in three simulations, where $\langle\rangle$ averages over all test particles.
In each simulation, $\sgrid=10^{-3}$ AU and $N_{\rm tp}=10^4$. Particles were initially on orbits evenly 
spaced in semi-major axis between $\bar{r}-\Delta_0/2$ and $\bar{r}+\Delta_0/2$, where
$\bar{r}= 1$ AU and $\Delta_0=0.08$ AU (top hat profile). Initial eccentricities were identical,
and initial longitudes and pericentre longitudes were random.

%%%%%%%%%%%%%%%%%%%%%%%%%%%%%%%%%%%%%%%%
\vbox{
\centerline{
\hglue+.2cm{\includegraphics[width=9.5cm]{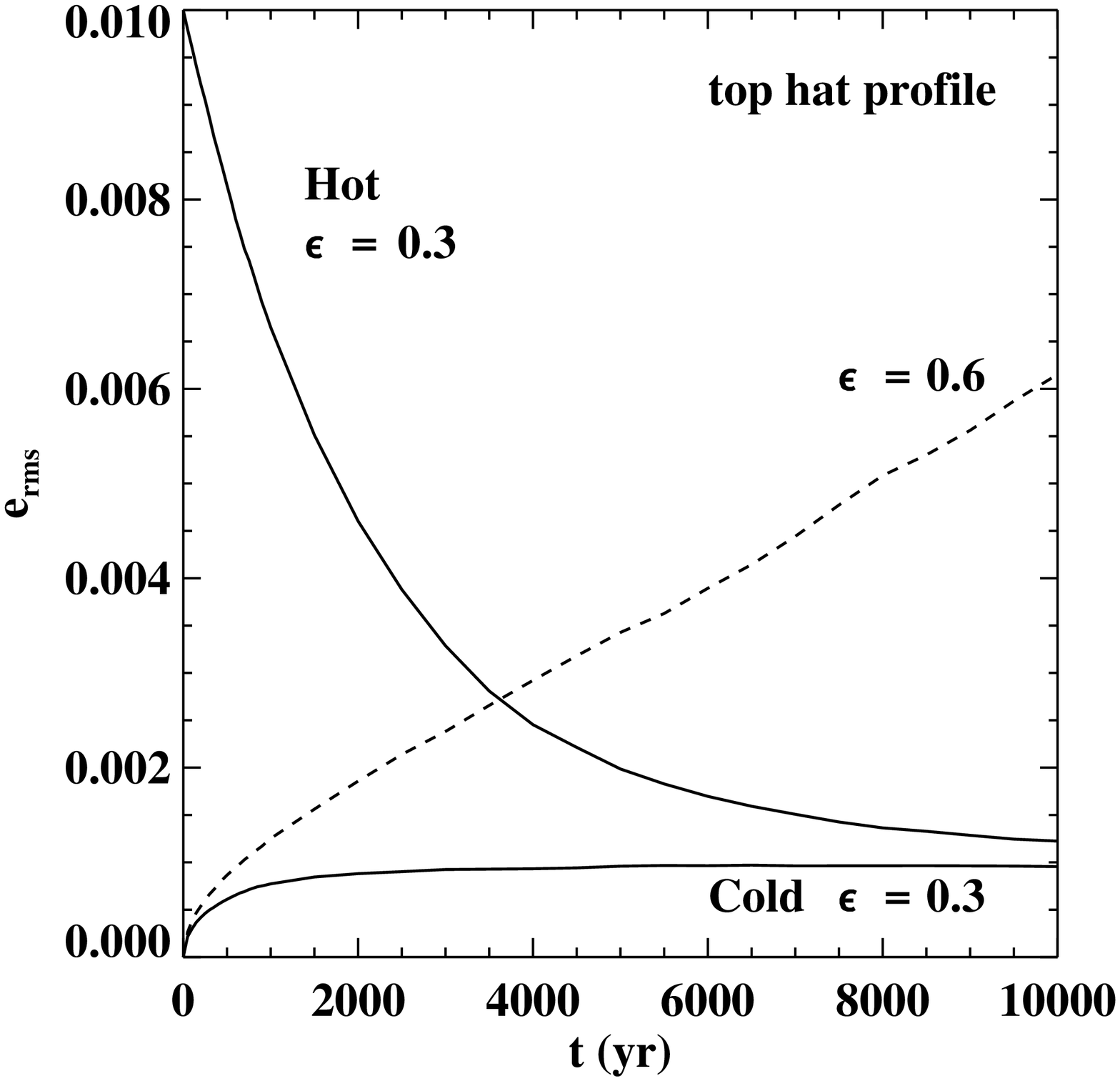}}
}
%\vspace{-.5cm}
\figcaption{Random Velocity Evolution.
The two simulations with $\epsilon=0.3$ both relax to the same $e_{\rm rms} \sim s_{\rm grid}/\bar{r}$.
In the $\epsilon=0.6$ simulation, heating proceeds indefinitely.
\label{fig:e_vs_t}}
}
%%%%%%%%%%%%%%%%%%%%%%%%%%%%%%%%%%%%%%%%

Two simulations had 
$\epsilon=0.3$. One of these was initially cold, with initial eccentricities $=0$;
the other was initially hot, with eccentricities $=0.01$.
In both simulations, $e_{\rm rms}$ approached $\sim s_{\rm grid}/\bar{r}=10^{-3}$ (Eq. [\ref{eq:emin}]).
Clearly, inelastic damping dominates viscous stirring when $\epsilon=0.3$.
The third simulation had $\epsilon=0.6$, and was initially cold.  Its $e_{\rm rms}$ grew indefinitely.
Hence viscous stirring dominates when $\epsilon=0.6$.

We define the collision time as
\be
t_{\rm col}\equiv {N_{\rm tp}/2 \over {\rm no.\ collisions\ per\ unit\ time}} \,.
\label{eq:coltime}
\ee
In the three simulations, we measured $t_{\rm col}=106-108$ yr.
An estimator of $t_{\rm col}$ from the input parameters of a simulation follows
from   a more precise
form of Equations (\ref{eq:taunew})-(\ref{eq:tcolnew}):
\be
t_{\rm col}^{\rm (est)}=t_{\rm orb}
{4\pi r\over ns_{\rm grid}^2} \ ,
\label{eq:tcolest2}
\ee
where 
\be
n\equiv dN_{\rm tp}/dr
\ee
 is the number of particles per radial distance, and $4\pi$ is the product of two factors: $2\pi$ for the area of a ringlet ($2\pi r \times dr$), and 2 because
only half of the time, when the relative velocity is negative, do doubly-occupied grid cells lead to collisions.
For the parameters of the present simulations, with
$r=\bar{r}$ and $dN_{\rm tp}/dr=N_{\rm tp}/\Delta_0$,
$t_{\rm col}^{\rm (est)}=101 $ yr.  
 Because
 $t_{\rm col}^{\rm (est)}$ does not account for the decreased collision frequency of particles at the
 edge of the ring, it underestimates $t_{\rm col}$ by 
 a small amount.   That $t_{\rm col}$ remained nearly constant throughout the simulation reflects
 the near constancy of the ring's width, since parameters were deliberately chosen to freeze out
 diffusion (from Eq. [\ref{eq:tdiff}] below, the diffusion time in these simulations is $\sim 10^5$ yr).

In the cold $\epsilon=0.3$ simulation, $e_{\rm rms}$ reached 80 percent of its final value by time $t = 1000$ yr---around
10 collision times.   
The hot $\epsilon=0.3$
simulation took much longer to reach velocity equilibrium:
initially, $e_{\rm rms}$ decayed approximately exponentially
with a time constant of $2500$ yr---around 25 collision times. 
Even after $t = 2500$ yr, the hot simulation
took hundreds of collision times to reach its final $e_{\rm rms}$.
Velocity equilibration in the hot simulation was so long
because of particles at ring edges.
{Edge particles tend to retain their initial eccentricities because their epicyclic excursions carry them away from the majority of particles; consequently, edge particles collide less frequently than do particles in the ring proper.}

%%%%%%%%%%%%%%%%%%%%%%%%%%%%%%%%%%%%%%%%
\vbox{
\centerline{
\hglue+.2cm{\includegraphics[width=9.5cm]{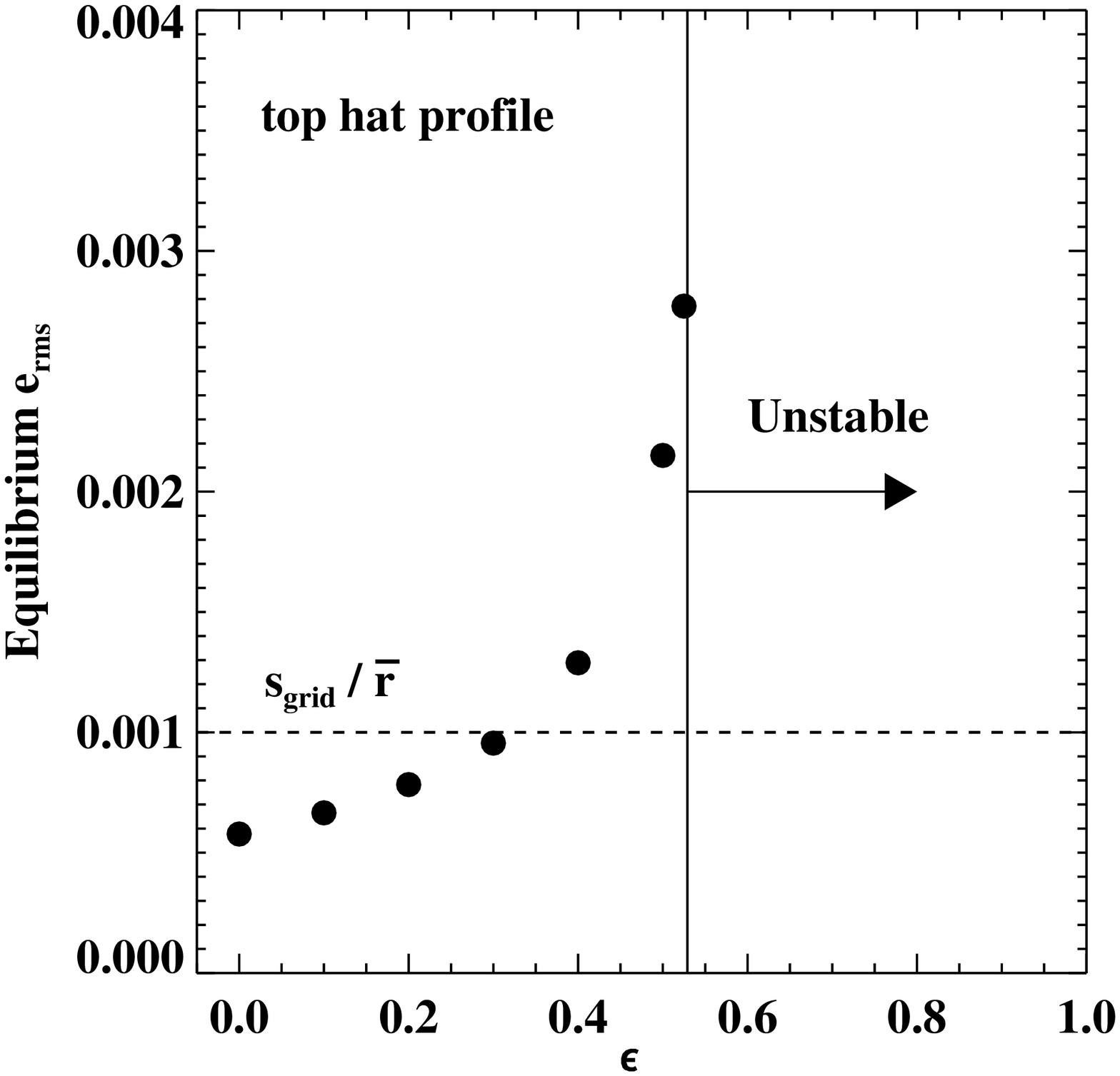}}
}
%\vspace{-.5cm}
\figcaption{
Equilibrium $e_{\rm rms}$.  Seven points denote the steady-state $e_{\rm rms}$
in simulations with differing $\epsilon$.
Vertical line shows critical value $\epsilon_*=\sqrt{7}/5=0.529$.
  In initially cold
simulations with $\epsilon=\{0,0.1,0.2,0.3\}$, 
$e_{\rm rms}$ was evaluated at $10^4$ yr. In initially cold
simulations with $\epsilon=\{0.4,0.5,0.525\}$,
$e_{\rm rms}$ was evaluated at $2\times 10^4$ yr, since these simulations took longer to reach steady state.
\label{fig:e_rms}}
}
%%%%%%%%%%%%%%%%%%%%%%%%%%%%%%%%%%%%%%%%

\subsection{Simulations: Equilibrium Eccentricities}
\label{sec:veleq}

Figure \ref{fig:e_rms} shows results from seven simulations, all with the same initial conditions
as the cold simulations described in the previous subsection, except with differing values of
$\epsilon$.  The seven plotted simulations, with $\epsilon \leq 0.525$, all reached velocity
equilibrium, with $e_{\rm rms}\sim {\rm (order\ unity\ constant)}\times s_{\rm grid}/\bar{r}$.  Simulations with $\epsilon\geq 0.6$ never reached velocity equilibrium.  We
conclude that $0.525<\epsilon_*<0.6$ for our collisional cross-section.  
In Equation (\ref{eq:star}) below, we derive $\epsilon_*=\sqrt{7}/5=0.529$.

\section{Density Evolution}
\subsection{Theory}
\label{sec:densitytheory}

A ring diffuses 
 in the time that it takes a particle to random walk across its width.
This random walk has a step-size equal to the epicyclic excursion of a particle ($\sim r e_{\rm rms}\sim s_{\rm grid}$)
and a time per random step
of $t_{\rm col}$.  So to diffuse the width of the ring $\Delta$ takes a time
\be
t_{\rm diff}\sim t_{\rm col}\left({\Delta \over s_{\rm grid}}\right)^2 \gg t_{\rm col} \,,
\label{eq:tdiff}
\ee
where the
inequality holds when $\Delta \gg s_{\rm grid}$; otherwise, $t_{\rm diff} \sim t_{\rm col}$ until $\Delta \sim s_{\rm grid}$.
Since $t_{\rm col}\propto 1/n\propto \Delta $ (Eq. [\ref{eq:tcolest2}]),
a ring  expands as
\be
\Delta \propto t^{1/3} \ .
\ee

More precisely, $n$
satisfies the diffusion equation
\begin{eqnarray}
{\partial n\over\partial t} &=&{\partial\over\partial r}\left(\nu {\partial n\over\partial r}\right) \ ,
\label{eq:dndt} 
\end{eqnarray}
where the viscosity $\nu = $ constant $\times \, s_{\rm grid}^2/t_{\rm col}$ (Eq. [\ref{eq:tdiff}]). Inserting Equation (\ref{eq:tcolest2}) into our
expression for $\nu$, we see that
\be
\nu= k_\nu{  s_{\rm grid}^4\over \bar{r} t_{\rm orb}} n \ ,
\label{eq:nu}
\ee
which defines the dimensionless constant $k_\nu$;
$k_\nu$ is a function of $\epsilon$, but is independent of $s_{\rm grid}$, $\bar{r}$, and
$n$.
 \cite{PH87} derived the above diffusion equation
and gave its self-similar solution, an expanding inverted parabola:
\be
n = {3\over 2}{N_{\rm tp}\over {\Delta }}\left(
1-\left({r-\bar{r}\over {\Delta }/2}\right)^2\right) \ , \ \ \vert r-\bar{r}\vert \leq {\Delta }/2
\label{eq:na}
\ee
where
\be
{\Delta  }\equiv 
\left(
36 k_\nu{  s_{\rm grid}^4\over  \bar{r} t_{\rm orb}}
 N_{\rm tp} t
\right)^{1/3} \ .
\label{eq:ahat}
\ee
Since the viscosity $\nu$ decreases with decreasing $n$, the diffusion is non-linear, and the edges of the ring  
at $r=\bar{r}\pm {\Delta  }/2$ are sharp.

\subsection{Simulations}
\label{sec:denssim}

Because of the
steep dependence of the diffusion timescale on the width of the ring, $t \propto \Delta^3$, it takes a long time
to simulate even a modest increase in $\Delta $.  Simulation parameters must be chosen judiciously.  We fix $N_{\rm tp}=10^4$ and $\bar{r}=1$ AU, and seek the optimal values for $s_{\rm grid}$
and
$\Delta_0\equiv \Delta \vert_{t=0}$.
To simulate as large an increase in $\Delta $ as possible, the simulation should begin
with as narrow a ring as possible.  For fixed $s_{\rm grid}$, the narrowest
ring that is not optically thick has unity optical depth: $\Delta_0 \sim N_{\rm tp}(s_{\rm grid})^2/2\pi \bar{r}$.
The evolution timescale at the start of the simulation is $t_0=$ constant $\times \;(\Delta_0)^3/(s_{\rm grid})^4$ 
(Eq. [\ref{eq:ahat}]), so with unity optical depth, $t_0=$ constant $\times \;\Delta_0$.  Hence the fastest
timescale is obtained with the smallest $\Delta_0$. But we must have
$\Delta_0\geq s_{\rm grid}$, so the optimal
values are
$\Delta_0=s_{\rm grid}=2\pi \bar{r}/N_{\rm tp}$.   Rounding up, we set
$\Delta_0=s_{\rm grid}=10^{-3}$ AU.

Figures \ref{fig:tcol+sigma_r} and \ref{fig:parabola} portray results from a simulation with the parameters listed above, and 
$\epsilon=0.3$. Initially, the ring particles were uniformly distributed in a ring with edges at 
$1\pm (5\times 10^{-4})$ AU. 
In Figure \ref{fig:tcol+sigma_r}, we show how the particles' dispersion in $r$,
\be
\sigma_r \equiv \langle (r-\bar{r})^2 \rangle^{1/2} \,,
\label{eq:sigmardef}
\ee
and how the collision time ($t_{\rm col}$) vary with time, according to both theory
and simulation. Figure \ref{fig:parabola} displays comparisons between theory and simulation for $n$.
According to the theory described in \S\ref{sec:densitytheory},
$n(r)$ is given by Equation
(\ref{eq:na}), the dispersion in $r$ is
\be
\sigma_r^2=\int (n/N_{\rm tp})(r-\bar{r})^2dr=\Delta^2/20
\label{eq:sigmahat}
\ee
with $\Delta$ given by Equation (\ref{eq:ahat}), and the collision time 
is
\be
t_{\rm col}= \left(\int {1\over N_{\rm tp}}{n\over t_{\rm col}^{\rm (est)}}dr\right)^{-1}=
{120\pi}k_\nu  {(s_{\rm grid})^2\over {\Delta }^2}t \ .
\label{eq:tcol}
\ee
The theory
fits the numerical results well when we set 
\be
k_\nu=0.016 \,.\label{eq:knureal}
\ee

%%%%%%%%%%%%%%%%%%%%%%%%%%%%%%%%%%%%%%%%
\vbox{
\centerline{
\hglue+.2cm{\includegraphics[width=9.5cm]{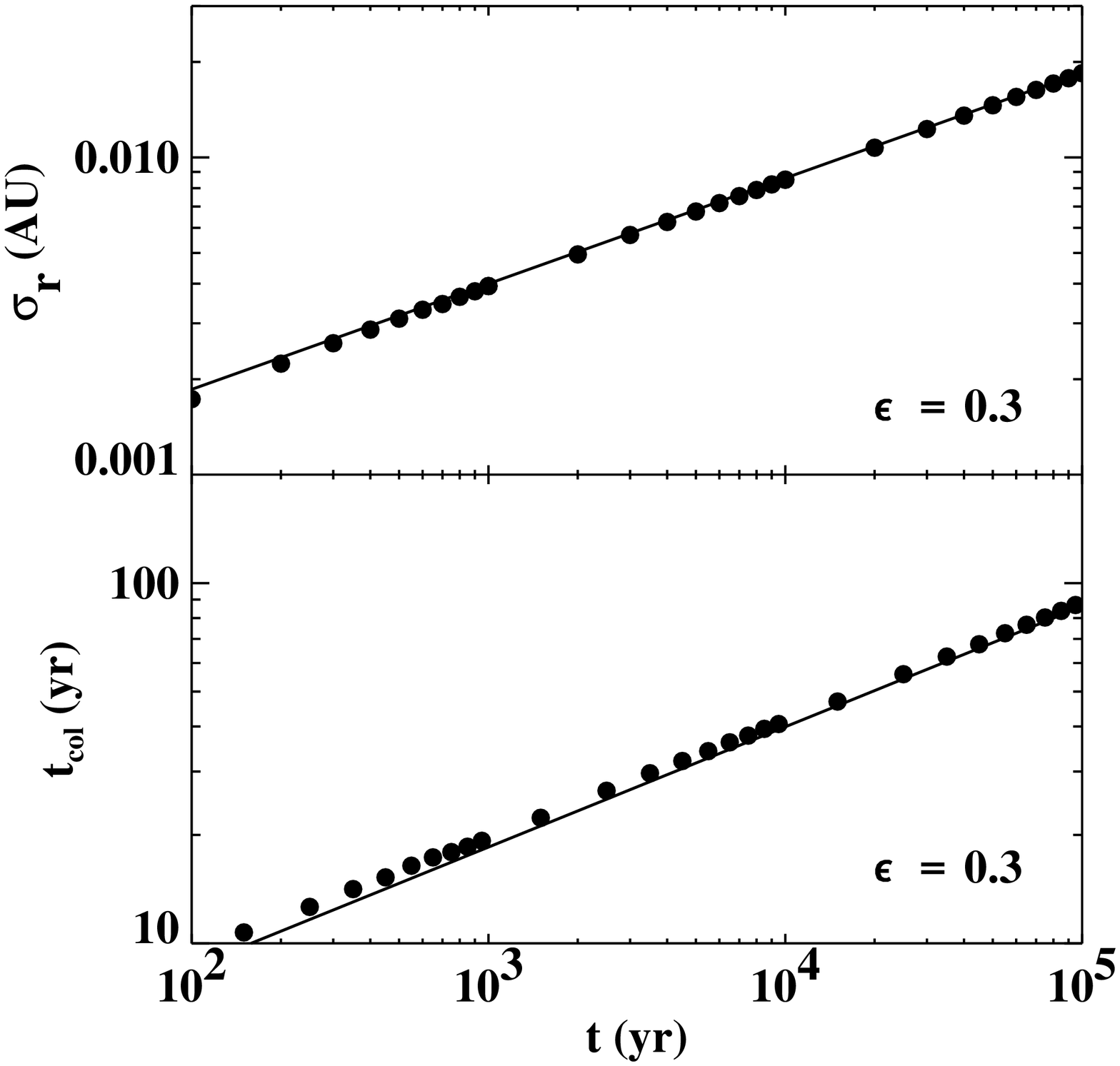}}
}
\figcaption{
Dispersion and Collision Rate Evolution in a Diffusing Ring.
Top panel: points show $\sigma_r=\langle(r-\bar{r})^2\rangle^{1/2}$ from the numerical simulation.
The theory line through the points is $\Delta /\sqrt{20}$ (Eq. [\ref{eq:sigmahat}]), where
 $\Delta $ is given by Equation (\ref{eq:ahat}).
 The normalization of the line was adjusted by choosing $k_\nu=0.016$.
Bottom panel: points show the collision time $t_{\rm col}$ from the simulation (Eq. [\ref{eq:coltime}]). The theory line is given
by Equation (\ref{eq:tcol}), with $k_\nu=0.016$. Theory underestimates $t_{\rm col}$ by a small 
amount because of particles at ring edges (\S \ref{sec:approach}).
\label{fig:tcol+sigma_r}}
}
%%%%%%%%%%%%%%%%%%%%%%%%%%%%%%%%%%%%%%%%
%%%%%%%%%%%%%%%%%%%%%%%%%%%%%%%%%%%%%%%%
\vbox{
\centerline{
\hglue+.2cm{\includegraphics[width=9.5cm]{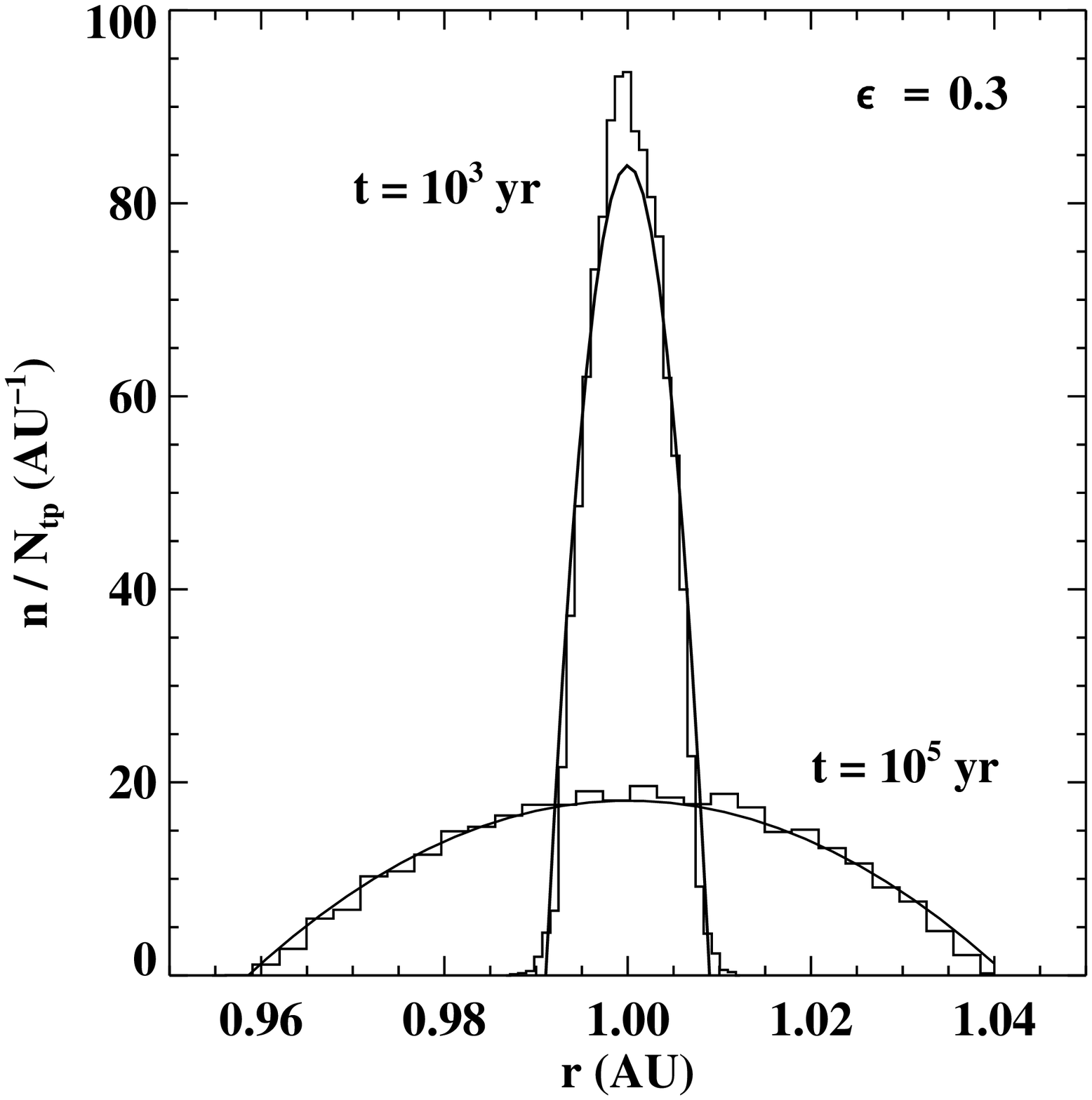}}
}
%\vspace{-.5cm}
\figcaption{Density Evolution in a Diffusing Ring. Histograms of the number
density are shown at two times from the same simulation as in
 Figure \ref{fig:tcol+sigma_r}.
 The theory lines through the histograms are given by 
Equation (\ref{eq:na}), with $k_\nu$ the same as in Figure \ref{fig:tcol+sigma_r}.
\label{fig:parabola}}
}
%%%%%%%%%%%%%%%%%%%%%%%%%%%%%%%%%%%%%%%%

\section{Angular Momentum and Energy Transport}

Understanding how planets interact with disks requires
understanding how angular momentum and energy are transported---both within
disks
and between planets and disks. Below we study transport in isolated,
circular, narrow rings, developing diagnostics that will prove useful
in future simulations of disks with planets.

\subsection{Theory}
\label{sec:amtheory}

The azimuthally-averaged equations describing the conservation of particle number, angular momentum, and
energy are
\begin{eqnarray}
\partial_t n & + & \partial_r F_n = 0 \label{eq:cont} \\
\partial_t (H n) & + & \partial_r (H F_n) = -\partial_r F_H^{\rm visc} \label{eq:am}\\
\partial_t (E n) & + & \partial_r (E F_n) = -\partial_r F_E^{\rm visc} - n\edot \ , \label{eq:e}
\end{eqnarray}
where $F_n$ is the (net) number flux across a circle of radius $r$, i.e., 
 the number
of particles per unit time that exit this circle minus the number that enter it; $H(r)\equiv ({GM_\odot r})^{1/2}$ 
is the specific angular momentum of a particle on a circular orbit; $E(r)\equiv -GM_\odot/2r$ is
the specific energy; $F_H^{\rm visc}$ (the ``viscous flux of angular
momentum'')
is defined as the difference
between the total  angular momentum flux and $HF_n$, i.e., $F_H^{\rm visc}+HF_n=F_H^{\rm tot}$;
$F_E^{\rm visc}$ is the corresponding difference in energy fluxes, i.e.,
$F_E^{\rm visc}+EF_n=F_E^{\rm tot}$; and $\edot$ is the rate
at which specific energy
is lost, per particle, in inelastic collisions.  

Our decomposition of $F_H^{\rm tot}$
into two components has the following interpretation:
$HF_n$ is the angular momentum flux that would be carried by particles
on circular orbits whose radii change on timescales long compared
to the orbit time, while
$F_H^{\rm visc}$ is the part of the angular momentum flux not associated with
the direct advection of circular orbits.  The viscous flux of angular
momentum and the corresponding viscous flux of energy
are transferred in the 
 ratio
appropriate for circular orbits, i.e., since circular orbits have 
$E=-(1/2)(GM_\odot/H)^2$, a transfer  in angular momentum of $\delta H$ must
be accompanied by a transfer in energy of $\delta E=(GM_\odot)^2 \delta H/H^3=\Omega\cdot \delta H$,
so
\be
{F_E^{\rm visc}\over F_H^{\rm visc}}=\Omega(r)\ .
\label{eq:viscratio}
\ee
The above relation applies only to the viscous fluxes, not to the total fluxes
(i.e., $EF_n/HF_n=-\Omega/2$).
Equations (\ref{eq:am}) and (\ref{eq:e}) simplify with the aid of Equations (\ref{eq:cont}) and (\ref{eq:viscratio}) to
\begin{eqnarray}
F_n&=&-{2\over r\Omega}\partial_r F_H^{\rm visc} \label{eq:fnfinal} \label{eq:lp74_a} \\
F_H^{\rm visc}&=&-n\edot/(d\Omega/dr)\ . \label{eq:lp74_b}
\end{eqnarray}
The latter is the well-known relation between energy dissipation and viscous angular momentum flux for
accretion disks \citep[e.g., ][]{LP74}.

Since $\edot \sim s_{\rm grid}^2\Omega^2/t_{\rm col} \sim ns_{\rm grid}^4\Omega^2/4\pi rt_{\rm orb}$, we may re-write (\ref{eq:cont}) using
(\ref{eq:lp74_a}) and (\ref{eq:lp74_b}) as
\be
\partial_t n=\left(\frac{k_Es_{\rm grid}^4}{2rt_{\rm orb}}\right)\partial_r^2 n^2 \,,
\label{eq:ke_full}
\ee
where
\be
k_E\equiv  {n\edot\over n^2 s_{\rm grid}^4 (3\Omega^2/8r t_{\rm orb})}
\label{eq:ke}
\ee
is an $\epsilon$-dependent dimensionless constant.
In deriving (\ref{eq:ke_full}), we have dropped terms that are small for
narrow rings (e.g., $|d(\ln \Omega)/dr| \ll |d(\ln n^2)/dr)|$).
Equation (\ref{eq:ke_full}) is identical to Equations
(\ref{eq:dndt})--(\ref{eq:nu}) provided $k_E=k_\nu$.

\subsection{Simulations}

In this subsection, we diagnose angular momentum and energy transport in numerical simulations and compare to theory.
To measure the number flux $F_n$ across a circle of radius $r$,
we evaluate $N_{>r}$,
the number of particles whose radial distances from the Sun exceed $r$,
at two times, $t_m$ and $t_m+dt_m$. Then
\be
F_n(r,t_m)\doteq{N_{>r}(t_m+dt_m)-N_{>r}(t_m)\over dt_m} \ , \label{eq:fnmeasure}
\ee
where the symbol $\doteq$ means  that this is how $F_n$ is measured.
Similarly, we measure the total angular momentum flux as
\be
F_H^{\rm tot}(r,t_m)\doteq{H_{>r}(t_m+dt_m)-H_{>r}(t_m)\over dt_m} \ ,
\ee
where $H_{>r}$ is the sum of the specific angular momenta of all particles
whose radial distances exceed $r$.
For the energy flux, we must account
for the energy lost in inelastic collisions
(where the specific energy lost per collision
$=u_{\rm rel,\parallel}^2(1-\epsilon^2)/4$; see Eq. [\ref{eq:upar}]).
To this end,
when calculating the energy flux across a circle with radius $r$, we first calculate
$\delta {\cal E}_{>r}$, the total specific
energy lost between times $t_m$ and $t_m+dt_m$ in all inelastic collisions that occurred at radii $>r$.
Then the total energy flux is
\be
F_E^{\rm tot}(r,t_m)\doteq{E_{>r}(t_m+dt_m)-E_{>r}(t_m)+\delta {\cal E}_{>r}\over dt_m} \ ,
\ee
where $E_{>r}$ is the sum of the specific energies of particles with radii greater than $r$.
The viscous fluxes are determined by the three fluxes above:
\begin{eqnarray}
F_H^{\rm visc}&\doteq&F_H^{\rm tot}-HF_n \label{eq:vm}\\
F_E^{\rm visc}&\doteq&F_E^{\rm tot}-EF_n \ . \label{eq:vm2}
\end{eqnarray}
Energy dissipation in an annulus between radius
$r_1$ and $r_2$ is measured via
 \be
 n\edot\doteq{\delta{\cal E}_{>r_1}-\delta{\cal E}_{>r_2}\over (r_2-r_1)dt_m} \ . \label{eq:em}
 \ee

Figure \ref{fig:fluxes} shows a number of measurements of the
 angular momentum flux for the simulation whose parameters are given
in \S \ref{sec:denssim}.
  The theory curve is from Equations
(\ref{eq:lp74_b}) and (\ref{eq:ke}):
\be
F_H^{\rm visc}={k_E}{n^2s_{\rm grid}^4\Omega\over 4 t_{\rm orb}} \ \label{eq:fhvisctheory} \ ,
\ee
with $n$ given by Equation (\ref{eq:na}), and
$k_E=k_\nu=0.016$ (we verify the equality of $k_E$ and $k_{\nu}$ in
Figure \ref{fig:k_e} below).
  Overlaid on this theory curve are three sets of datapoints, measured
 in three independent ways.  The agreement between theory and simulations is
excellent.

Figure \ref{fig:f_n} shows the number fluxes for the same simulation.
The theory line shows (Eqs. [\ref{eq:fnfinal}],[{\ref{eq:fhvisctheory}}])
\be
F_n=-k_E {s_{\rm grid}^4\over 2\bar{r} t_{\rm orb}}{\partial n^2\over \partial r} \ , \label{eq:fntheory}
\ee
with $n$ from Equation (\ref{eq:na}).
The datapoints agree well with the theory, although there is some scatter.

The circles in Figure \ref{fig:k_e} show the energy dissipation constant
$k_E$ (Eq. [\ref{eq:ke}])
 for the simulations described in 
\S\S \ref{sec:approach}--\ref{sec:veleq}, at the same output times (Fig.  \ref{fig:e_rms}). Recall that each of these simulations has a top hat
density profile and is run for much less than a viscous time, so the density
hardly evolves.   The simulation with $\epsilon=0.3$ has $k_E=0.016$.  Therefore
$k_E=k_\nu$ (Eq. [\ref{eq:knureal}]), as suggested
below Equation (\ref{eq:ke}).  
Note that it is much more efficient to measure $k_E$ than $k_\nu$, since $k_E$ can 
be measured in only a few collision times, whereas $k_\nu$ must be measured on the viscous timescale.
We defer to \S \ref{sec:simlast}
a discussion of the diamonds and stars in Figure \ref{fig:k_e}.

%%%%%%%%%%%%%%%%%%%%%%%%%%%%%%%%%%%%%%%%
\vbox{
\centerline{
\hglue+.2cm{\includegraphics[width=9.5cm]{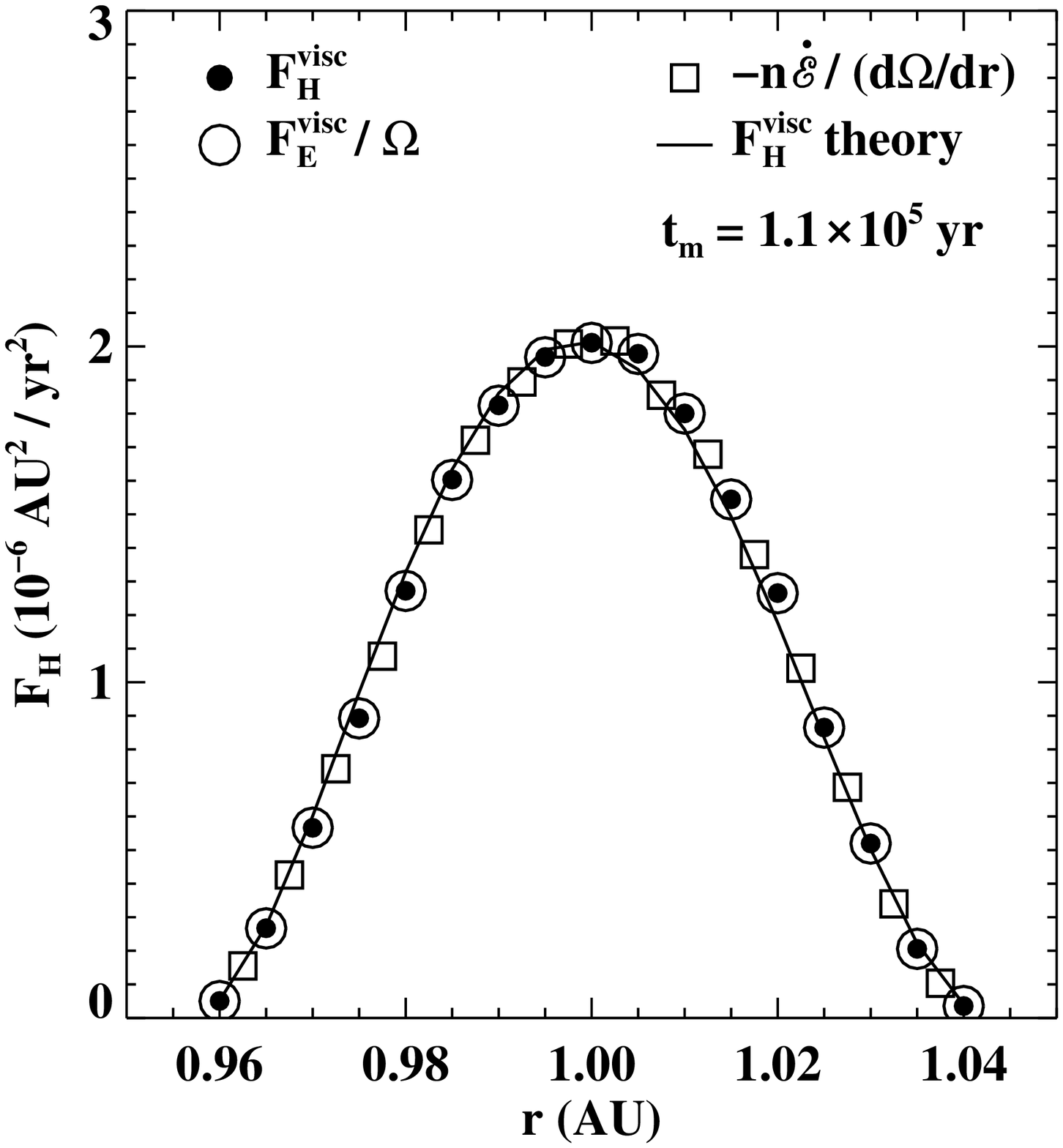}}
}
%\vspace{-.5cm}
\figcaption{
Angular Momentum Flux Measured With Various Methods. Data are taken from 
the simulation described in \S \ref{sec:denssim} (Figs. \ref{fig:tcol+sigma_r}--\ref{fig:parabola}) at
time 
$t_m=1.1\times 10^5$ yr, with measurement interval $dt_m=10^4$ yr.
The theory line is Equation (\ref{eq:fhvisctheory}).
 $F_H^{\rm visc}$ is measured with Equation (\ref{eq:vm}).
     $F_E^{\rm visc}$ is measured with Equation (\ref{eq:vm2}); clearly      $F_E^{\rm visc}/\Omega=F_H^{\rm visc}$, confirming Equation (\ref{eq:viscratio}).
   $n\edot$ is measured with Equation (\ref{eq:em}); the squares confirm Equation (\ref{eq:lp74_b}).
\label{fig:fluxes}}
}
%%%%%%%%%%%%%%%%%%%%%%%%%%%%%%%%%%%%%%%%

%%%%%%%%%%%%%%%%%%%%%%%%%%%%%%%%%%%%%%%%
\vbox{
\centerline{
\hglue+.2cm{\includegraphics[width=9.5cm]{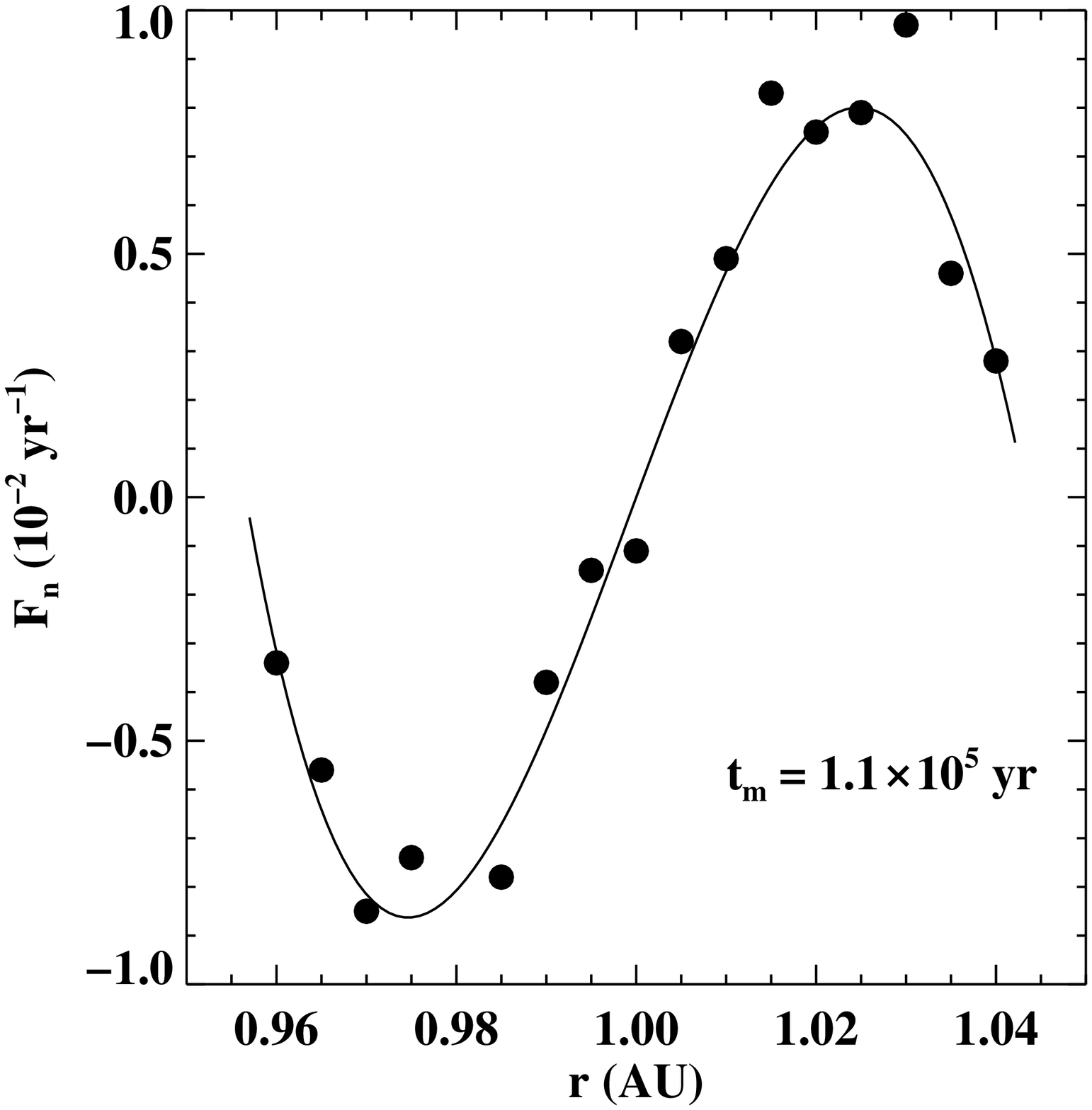}}
}
%\vspace{-.5cm} 
\figcaption{
Number Flux. Data are taken from the same simulation as in Figure \ref{fig:fluxes}, and at same times.
The theory line is Equation (\ref{eq:fntheory}). Datapoints
are measured with Equation (\ref{eq:fnmeasure}).
\label{fig:f_n}}
}
%%%%%%%%%%%%%%%%%%%%%%%%%%%%%%%%%%%%%%%%
 
%%%%%%%%%%%%%%%%%%%%%%%%%%%%%%%%%%%%%%%%
\vbox{
\centerline{
\hglue+.2cm{\includegraphics[width=9.5cm]{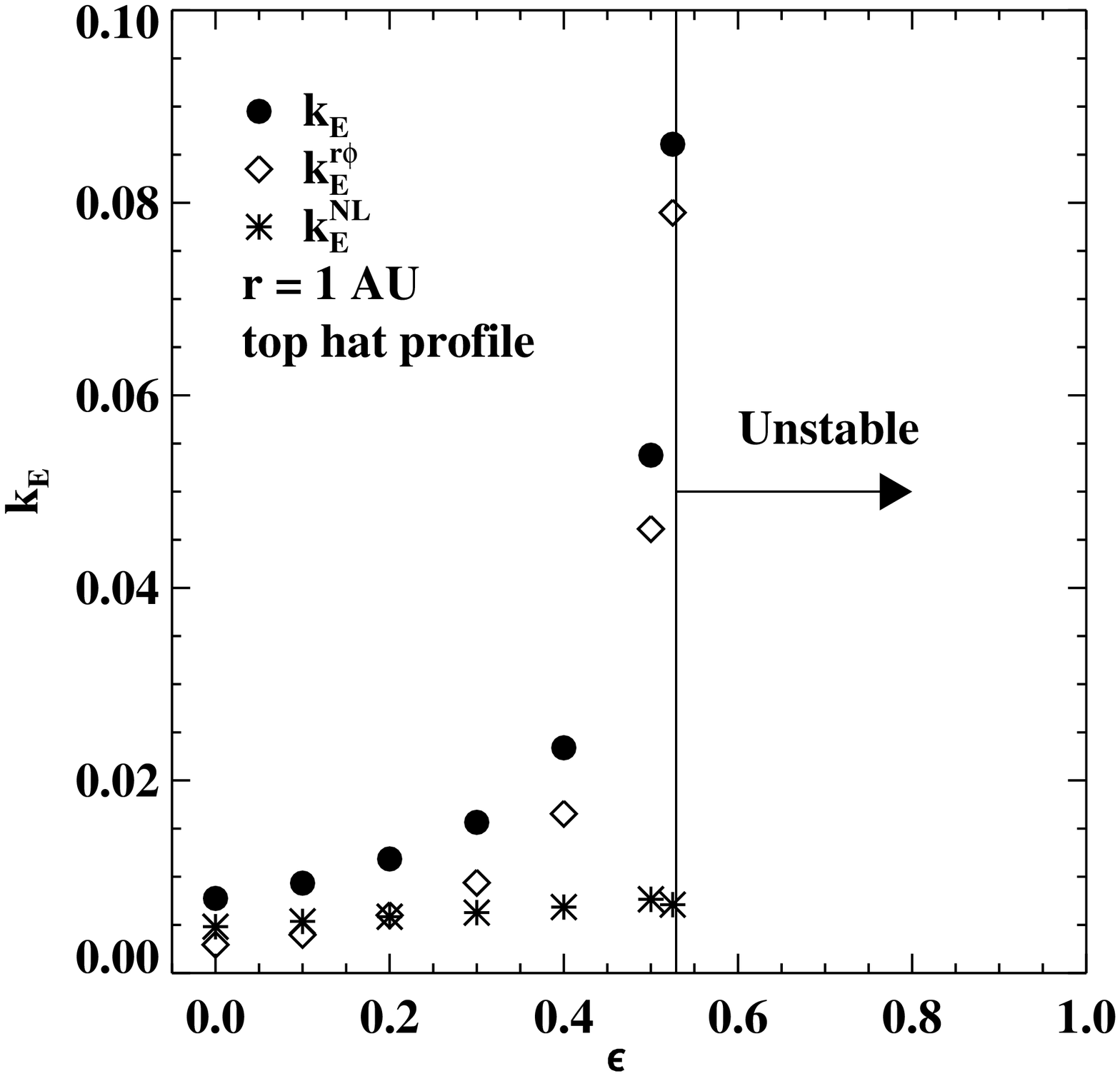}}
}
%\vspace{-.5cm}
\figcaption{
Energy Dissipation.  The simulations are the same as those described in Figure \ref{fig:e_rms}. Circles show $k_E$, measured with Equations (\ref{eq:ke}) and (\ref{eq:em}). Measurement parameters include $r_1 = 0.99$ AU, $r_2 = 1.01$
AU, $dt_m = 5000$ yr, and $t_m = 5000$ yr for
$\epsilon=\{0,0.1,0.2,0.3\}$ and $t_m = 15000$ yr
for $\epsilon=\{0.4,0.5,0.525\}$.
For $\epsilon = 0.3$, $k_E = 0.016$, a value that matches $k_{\nu}$ as given
by Equation (\ref{eq:knureal}).
Diamonds show $k_E^{r\phi} \propto F_{r\phi}$,
measured as described in \S\ref{sec:simlast}.
Stars give $k_E^{\rm NL} = k_E - k_E^{r\phi} \propto F_{\rm NL}$;
this quantity is practically constant with $\epsilon$,
as expected from Equation (\ref{eq:fb}).
Most of the energy dissipation (viscous transport of angular momentum)
arises from $F_{r\phi}$
and not from $F_{\rm NL}$ as $\epsilon$ approaches $\epsilon_*$; compare
with Equations (\ref{eq:fa})--(\ref{eq:fb}).
\label{fig:k_e}}
}
%%%%%%%%%%%%%%%%%%%%%%%%%%%%%%%%%%%%%%%%

\subsection{Dynamics from a Microscopic Perspective: Theory}
\label{sec:micro_theory}

The angular momentum flux advected by particles
is 
\begin{eqnarray}
&=&n\langle r(\Omega r+v_\phi)v_r\rangle \\
&=& HF_n +nr\langle v_rv_\phi\rangle \,,
\end{eqnarray}
where $\bld{v}=(v_r,v_\phi)$  is the difference between a particle's total velocity (in the radial $r$ and azimuthal $\phi$ directions)
and the Keplerian circular velocity at its position, and $\langle\rangle$ denotes
an average over particles in a narrow ring.
Defining
\be
F_{r\phi}\equiv nr\langle v_rv_\phi\rangle \,,
\ee
we have
\be
F_H^{\rm tot}=HF_n+F_{r\phi}+F_{\rm NL} \ ,
\ee
where $F_{\rm NL}$ is the ``non-local'' flux, i.e, the angular momentum flux not advected
by particles \citep{WT88}.   In \S \ref{sec:amtheory}, we considered only the combination
\be
F_H^{\rm visc}=F_{r\phi}+F_{\rm NL} \,.
\label{eq:combo}
\ee
In this subsection, we wish to calculate $F_H^{\rm visc}$ in terms of microscopic quantities.
Hence we must consider the two components of $F_H^{\rm visc}$ separately.
The two components  have
the following interpretation:
(i) 
A particle that crosses a circle of radius $r$ has an angular momentum $H'$
that is not, in general, equal to $H(r)$; the difference $H'-H(r)$ contributes to $F_{r\phi}$.
(ii) When a particle that is inside of the circle collides with one that is outside, the angular 
momentum transferred across the circle contributes to $F_{\rm NL}$.

For a collisionless Keplerian particle, to lowest order in $e$,
\begin{eqnarray}
v_r&=&er\Omega\sin(\Omega t) \label{eq:vr}\\
v_\phi&=&{1\over 2}er\Omega \cos(\Omega t) \  ,\label{eq:vphi}
\end{eqnarray}
with $r,\Omega$ independent of time.
Therefore for collisionless particles, $F_{r \phi} \propto \langle v_rv_\phi \rangle \propto \langle \sin(2\Omega t) \rangle = 0$.
But collisions give a definite contribution to $\langle v_rv_\phi\rangle$
\citep{Greenberg88},
as we presently show.  

Instead of averaging over space only (as denoted by $\langle v_rv_\phi\rangle$), it will prove convenient
to average over both space and time. We average
over a radial width $\Delta r$
that is larger than the particle size but
smaller than the lengthscale over which the density varies.
We also average over a time $\Delta t$ that is longer than
a few collision times but shorter than the timescale over which the
density changes. The average $\overline{F}_{r\phi}$ is defined via
\be
(\Delta t\Delta r)\overline{F}_{r\phi}\equiv
\int_{t}^{t+\Delta t}\int_{r}^{r+\Delta r}dt'dr' (nr'v_rv_\phi) \ , \label{eq:fav}
\ee
where $\bld{v}$ is a Lagrangian quantity that is tied to particles.
Extracting  $F_{r\phi}$ from a numerical simulation with only 
a spatial average  ($\langle v_rv_\phi\rangle$) can lead to large errors.
In particular, in an optically thin disk, only a small fraction
of the particles have collided within the last orbital period.
Hence most particles contribute negligibly to 
$\langle v_rv_\phi\rangle$, and a calculation of $\langle v_rv_\phi\rangle$ can become plagued by small-number statistics.

For a single particle that does not collide in the interval $\Delta t$,
\be
\int_{t}^{t+\Delta t} v_rv_\phi dt'=-\Omega^{-1}\int {d v_\phi^2\over dt}dt = -\Omega^{-1}
\left(v_\phi^2\vert_{t+\Delta t}-v_\phi^2\vert_t  \right) \,.
\ee
But if it collided once, 
\be
\int_t^{t+\Delta t}v_rv_\phi dt'=-\Omega^{-1}\left(
v_\phi^2\vert_{t+\Delta t}-v_\phi^2\vert_t  -\delta(v_\phi^2)
\right) \ ,
\ee
where $\delta$ represents the change due to the collision.
Therefore each collision 
contributes to
Equation (\ref{eq:fav}) in the amount  of
\be
(\Delta t\Delta r) \, \delta\overline{F}_{r\phi}=r\Omega^{-1}(\delta(v_{\phi,1}^2)+\delta(v_{\phi,2}^2)) \ ,
\label{eq:flocal}
\ee
where $\delta(v_{\phi,1}^2),\delta(v_{\phi,2}^2)$ are the
contributions from the two collision partners. The contribution from the
endpoints,
$v_\phi^2\vert_{t+\Delta t}-v_\phi^2\vert_t$, can be neglected as long as
$\Delta t$ is much longer than 
the collision time.

We define
\be
(\Delta t\Delta r)\overline{F}_{\rm NL}\equiv\int\int dt' dr' F_{\rm NL} \,.
\label{eq:integral_NL}
\ee
If two particles collide when their positions are at radii $r_1, r_2$
(where $r < r_2 < r_1 < r+\Delta r$), and
if the particle at $r_1$
has its $\phi$-velocity changed by $\delta v_{\phi,1}$
in the collision,
then the collision contributes to the integral in
Equation (\ref{eq:integral_NL}) in the amount of
\begin{eqnarray}
(\Delta t\Delta r)\,\delta \overline{F}_{\rm NL}
&=&(r_1-r_2)r_1 \delta v_{\phi,1} \label{eq:fnl} \\
&\approx&r(r_1-r_2)(\delta v_{\phi,1}-\delta v_{\phi,2})/2 \,. \label{eq:fnl2}
\end{eqnarray}
To obtain the latter symmetric form, we
approximated $r_1 \approx r$
and $\delta v_{\phi,1} \approx -\delta v_{\phi,2}$ (when in fact
$r_1 \delta v_{\phi,1} = -r_2 \delta v_{\phi,2}$). The error accrued
is of order $(r_1-r_2)/r \sim s/r \ll 1$, where $s$ is the particle size.

Since the number of collisions per unit time per unit radius is $n/2t_{\rm col}$,
\ba
\overline{F}_{r\phi}&=&{n\over 2t_{\rm col}}
{2r\over \Omega}
\langle\delta(v_\phi^2)\rangle_c  \ , \label{eq:f1} \\
\overline{F}_{\rm NL}&=&{n\over 2t_{\rm col}}{r\over 2}\langle(r_1-r_2)\delta (v_{\phi,1}-v_{\phi,2})\rangle_c \label{eq:f2} 
\ea
where $\langle\rangle_c$ (not to be confused with $\langle\rangle$) is an average over collisions,
and $\langle\delta(v_\phi^2)\rangle_c \equiv \langle\delta(v_{\phi,1}^2+v_{\phi,2}^2)\rangle_c/2$.
  With our collision algorithm,
 $t_{\rm col}$ may be pulled out of the averages. We use Equations (\ref{eq:f1}) and (\ref{eq:f2}) to extract $\overline{F}_{r\phi}$ and $\overline{F}_{\rm NL}$ from
the simulations (see Figure \ref{fig:micro} below).

We can estimate the magnitudes of the two fluxes as follows.
Since the peculiar velocity
distribution is anisotropic, with $\langle v_\phi^2\rangle=\langle v_r^2\rangle/4<\langle v_r^2 \rangle$ (Eqs. [\ref{eq:vr}]-[\ref{eq:vphi}]), 
and since collisions tend to isotropize the distribution,  therefore collisions systematically
increase $v_\phi^2$  by
$\langle \delta(v_\phi^2)\rangle_c\sim v^2\sim  +e_{\rm rms}^2(r\Omega)^2$,
 transporting of ${F}_{r\phi}$ outwards.
The contribution to ${F}_{\rm NL}$ is 
$\langle(r_1-r_2)\,\delta(v_{\phi,1}-v_{\phi,2})\rangle_c\sim + \Omega s^2$, where $s$ 
is the particle size.  This is true even when $e_{\rm rms}\gg s/r$ (i.e., when $\epsilon\rightarrow\epsilon_*$),
because in that case  $\delta (v_{\phi,1}-v_{\phi,2})$ is nearly random and hence nearly
uncorrelated with $r_1-r_2$.  But because of the mean Keplerian shear,
a small correlation $\sim \Omega s^2$ persists. This
contribution also transports angular momentum outward. In sum,
\ba
\overline{F}_{r\phi}&\sim&r\Omega{n\over t_{\rm col}} (re_{\rm rms})^2 \label{eq:fa} \\
\overline{F}_{\rm NL}&\sim&r\Omega{n\over t_{\rm col}} s^2  \ . \label{eq:fb}
\ea

\subsubsection{Relating $n\edot$ to $F_H^{\rm visc}$}

We now calculate the energy lost in an inelastic collision, and thereby
re-derive Equation (\ref{eq:lp74_b}) from a microscopic perspective. 
Consider the collision of two particles having
total velocities $\bld{V}_{1,2}=\bld{v}_{1,2}+\bld{V}^{\rm circ}_{1,2}$,
where $\bld{V}^{\rm circ}_{1,2}=\Omega(r_{1,2}) r_{1,2} \bld{\hat{\phi}}$ 
is the circular Keplerian speed at the positions of the particles.
Then the specific energy lost per collision is, in previous notation,
\begin{eqnarray}
(\Delta t\Delta r)\,\delta\left(
\overline{n\edot}
\right)
&=&
-
\delta(V_1^2+V_2^2)/2
\\
&=&
-
\delta(v_1^2+v_2^2)/2 \nonumber\\
& &-\delta\bld{v}_1
\bld{\cdot V}^{\rm circ}_{1}
-\delta\bld{v}_2
\bld{\cdot V}^{\rm circ}_{2}
\\
&=&
-\delta(v_1^2+v_2^2)/2 \nonumber \\
&&-(r_1-r_2)r_1\delta v_{\phi,1}d\Omega/dr \label{eq:edot_1} \\
& \approx & 
-\delta(v_1^2+v_2^2)/2 \nonumber \\
&&-(r_1-r_2)r (d\Omega/dr) \times \nonumber \\
&&(\delta v_{\phi,1} - \delta v_{\phi,2})/2 \,, \label{eq:edot_2}
\end{eqnarray}
where to derive (\ref{eq:edot_1}) we used conservation of orbital
angular momentum in a collision ($r_1 \delta v_{\phi,1} = -r_2 \delta v_{\phi,2}$). Therefore
\be
\overline{n\edot}=
-{n\over 2t_{\rm col}}
\langle
\delta(v^2)
+{r\over 2}{d\Omega\over dr}(r_1-r_2)\delta(v_{\phi,1}-v_{\phi,2}) 
\rangle_c \ . \label{eq:ne}
\ee
We re-write the first term using
   $v^2=-3v_\phi^2+e^2(r\Omega)^2$
 (Eqs. [\ref{eq:vr}]--[\ref{eq:vphi}]).
Since the particles are in collisional equilibrium, 
\be
\langle
\delta(e^2)
\rangle_c=0 \ , \label{eq:coleq}
\ee
which implies that $\langle \delta(v^2) \rangle_c=-3\langle\delta(v_\phi^2)\rangle_c$.
Inserting this into Equation (\ref{eq:ne}), and comparing the result with
 Equations (\ref{eq:f1})--(\ref{eq:f2}), completes our proof of Equation
 (\ref{eq:lp74_b}).

\subsubsection{Calculating $F_{r\phi}$ and $\epsilon_*$}

We calculate the numerical
constant that we dropped in Equation (\ref{eq:fa}), and
then use that result to calculate $\epsilon_*$.
\cite{GT78} 
 perform a similar calculation, with a different (though still idealized) collisional cross-section. \cite{Shukhman84} accounts for ${F_{\rm NL}}$ as well.
The treatments of \cite{GT78} and \cite{Shukhman84} are
more rigorous than ours, as they integrate over the velocity
distribution function.
 They also consider the more general case
of disks with order-unity optical depth.
But their final expressions are ``extremely
cumbersome'' \citep{Shukhman84}. 
Although our treatment is not rigorous, it is considerably simpler,
and we justify it by comparing with simulations.
We neglect ${F}_{\rm NL}$ in the present subsection, taking
$e_{\rm rms}\gg s/r$.

For $\overline{F}_{r\phi}$ (Eq. [\ref{eq:f1}]), we seek
\be
\langle
\delta(v_\phi^2)
\rangle_c=
\langle v_\phi^2 \rangle_{ac}-
\langle
v_\phi^2
\rangle_{bc} \ , 
\ee
where $\langle\rangle_{bc}$ is an average over collisions of the state immediately
before the collision, and $\langle\rangle_{ac}$ is of the state immediately after.
In an optically thin disk, the averages over particles are equal to time-averages for a single
particle:
\be
\langle
v_r^2
\rangle = {e_{\rm rms}^2\over 2} (r\Omega)^2 \ \ , \ \ \ 
\langle
v_\phi^2
\rangle =
{e_{\rm rms}^2\over 8}(r\Omega)^2 \label{eq:v1} \ ,
\ee
using Equations (\ref{eq:vr})--(\ref{eq:vphi}).
With our collision algorithm, the probability that a particle collides is uniform in time.
Therefore
\be
\langle v_{\{r,\phi\}}^2\rangle_{bc}=\langle v_{\{r,\phi\}}^2\rangle \ .
\ee 
In collisional equilibrium (Eq.  [\ref{eq:coleq}]),
\be
\langle
\delta(v_r^2+4v_\phi^2)
\rangle_c = 0 \ . \label{eq:vv}
\ee
We now evaluate $\langle v_r^2\rangle_{ac}$ and $\langle
v_\phi^2 \rangle_{ac}$.  We make the plausible assumption that the
relative velocity of collision 
partners,\footnote
{Since we take $e\gg s/r$, we neglect
the difference in the Keplerian circular velocities at the positions of the two particles.}
 ${\bf u}= {\bf v}_1-{\bf v}_2$,
is isotropically distributed after the collision:
\be
\langle u_r^2\rangle_{ac}=\langle u_\phi^2\rangle_{ac} \ . \label{eq:uu}
\ee
This assumption is verified by numerical simulation in \S\ref{sec:simlast}.
To relate ${\bf u}$ to ${\bf v}$,
 \ba
 \langle u_{\{r,\phi\}}^2\rangle_{bc}&=&2\langle v_{\{r,\phi\}}^2\rangle_{bc}  \\
  \langle \delta(u_{\{r,\phi\}}^2)\rangle_c&=&4\langle\delta(v_{\{r,\phi\}}^2)\rangle_c \ .
        \label{eq:ur}
 \ea
 For the first relation, we neglected
    the correlation 
 between $\bld{v}_1$ and $\bld{v}_2$ before a collision,
  and for  the second relation, we used  $\delta(\bld{v}_1+\bld{v}_2)=0$.
 Equations (\ref{eq:uu})--(\ref{eq:ur}) yield $\langle\delta(v_\phi^2-v_r^2)\rangle_c=(1/2)\langle v_r^2-v_\phi^2\rangle_{bc}$, which, with Equations (\ref{eq:v1})--(\ref{eq:vv}) becomes
 \be
 \langle
 \delta(v_\phi^2)
 \rangle_c={3\over 80}e_{\rm rms}^2 (r\Omega)^2 \ , \label{eq:ou}
\ee
giving the numerical constant for $\overline{F}_{r\phi}$ (Eq. [\ref{eq:f1}]).

To calculate $\epsilon_*$, we evaluate the specific energy lost per collision in the center-of-mass
frame:
\be
-{1\over 4}\langle\delta(u^2)\rangle_c={9\over 80}e_{\rm rms}^2(r\Omega)^2 \ , \label{eq:e1}
\ee
(Eqs [\ref{eq:vv}], [\ref{eq:ur}], [\ref{eq:ou}]).
We can also evaluate the energy loss as follows: from Equation (\ref{eq:upar}), it is
\be
{1\over 4}(1-\epsilon^2)\langle u_{\rm rel,\parallel}^2\rangle_{bc}=
(1-\epsilon^2){1\over 8}\langle u^2\rangle_{bc} \ , \label{eq:e2}
\ee
where we have made use of
 the isotropy of the pre-collision $\bld{u}$ with respect to the axis connecting the centers
 of the two particles.
Since $\langle u^2\rangle_{bc}=(5/4)e_{\rm rms}^2(r\Omega)^2$, Equations (\ref{eq:e1}) and
 (\ref{eq:e2}) are equal only if $\epsilon$ is equal to
\be
\epsilon_*= {\sqrt{7}\over 5}= 0.529 \ . \label{eq:star}
\ee
This is interpreted as the one value for $\epsilon$ that enables the
velocity distribution to equilibrate in the limit that
$e_{\rm rms} \gg s/r$, i.e., $F_{r\phi} \gg F_{\rm NL}$.

 \subsection{Dynamics from a Microscopic Perspective: Simulations}
 \label{sec:simlast}
 
In Figure \ref{fig:micro} we test the theoretical results
derived in \S\ref{sec:micro_theory}. In the top panel, for the
simulation with the same initial conditions
as the one with $\epsilon=0.525$ in Figures \ref{fig:e_rms} and \ref{fig:k_e},
we compute $\langle \delta (v_{\phi})^2 \rangle_c$---the key factor that
enters into $F_{r\phi}$ (Equation [\ref{eq:f1}])---in two ways: first as an
average over all particles that collide within successive time intervals
of 2000 yr, and second using Equation (\ref{eq:ou}), where
$e_{\rm rms}^2 = \langle e^2 \rangle$ is a spatial, not temporal, average
over all particles (regardless of whether they collide).
The agreement verifies Equation (\ref{eq:ou}).
Also shown in the top panel is the corresponding contribution to
$F_{\rm NL}$, i.e., $(\Omega/4)\langle(r_1-r_2)\delta(v_{\phi,1}-v_{\phi,2})\rangle_c$ (Equation [\ref{eq:f2}]). That this quantity is constant with time
and sits far below $\langle \delta (v_{\phi})^2 \rangle_c$ is expected
from Equations (\ref{eq:fa})--(\ref{eq:fb}), given this simulation in which
$e_{\rm rms} \gg s_{\rm grid}/\bar{r}$ (i.e.,
$\epsilon_* - \epsilon \ll \epsilon_*$).

In the bottom panel of Figure \ref{fig:micro}, we test our assumption, made in
Equation (\ref{eq:uu}), that post-collision relative velocities are isotropic.
The measured near-equality between $\langle u_r^2 \rangle_{ac}$
and $\langle u_\phi^2 \rangle_{ac}$ is satisfactory.

Finally, returning to Figure \ref{fig:k_e}, we plot separately the
contributions to the total energy dissipation from $F_{r\phi}$ and
$F_{\rm NL}$. The former contribution is described by
\be
k_E^{r\phi} \equiv  {-F_{r\phi} (d\Omega/dr) \over n^2 s_{\rm grid}^4 (3\Omega^2/8r t_{\rm orb})}
\label{eq:ke_rphi}
\ee
(see Equations [\ref{eq:lp74_b}], [\ref{eq:ke}], and [\ref{eq:combo}]).
We measure $F_{r\phi}$ according to
\be
\overline{F}_{r\phi} \doteq \left( \frac{r}{\Omega} \right) \frac{\delta (v_{\phi}^2)_{> r_1} - \delta (v_{\phi}^2)_{> r_2}}{dt_m (r_1 - r_2)} \,,
\ee
where $\delta(v_{\phi}^2)_{> r}$ is the total change in $v_{\phi}^2$
summed over particles that collide between times
$t_m$ and $t_m + dt_m$, at radii $> r$.
For simplicity, we evaluate $k_E^{\rm NL} = k_E - k_E^{r\phi}$.

%%%%%%%%%%%%%%%%%%%%%%%%%%%%%%%%%%%%%%%%
\vbox{
\centerline{
\hglue+.2cm{\includegraphics[width=9.5cm]{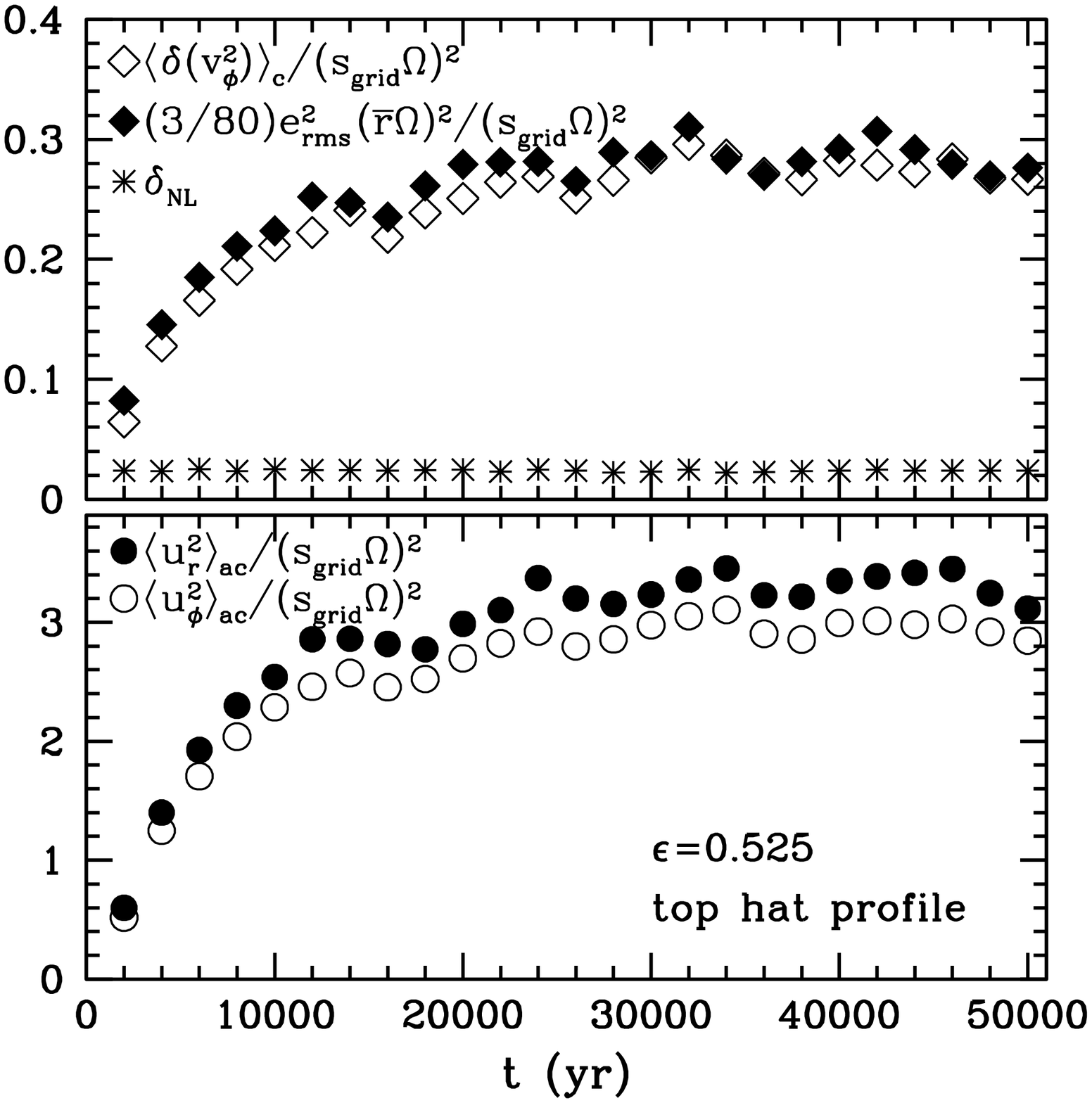}}
}
%\vspace{-.5cm}
\figcaption{Averages over Collisions.
The two panels show the time evolution of various collisional averages in a simulation 
with the same initial conditions as the one with $\epsilon=0.525$ in Figures \ref{fig:e_rms}
and \ref{fig:k_e}. Top panel: 
open diamonds show  the factor that enters into
 $F_{r\phi}$ (Eq. [\ref{eq:f1}]), i.e.,
$\langle\delta(v_\phi^2)\rangle_c=(1/2)\langle\delta(v_{\phi,1}^2+v_{\phi,2}^2)\rangle_c$, normalized as shown.  The averaging over collisions is done by recording
 the peculiar velocities of colliding particles immediately before and after each collision.
   The averaging time is $2000$ yr.
 Solid diamonds give $e_{\rm rms}^2=\langle e^2\rangle$ (a spatial, not temporal, average over all particles),
 multiplied by the appropriate pre-factor as given by Equation (\ref{eq:ou}). 
  The agreement between open and closed diamonds confirms Equation (\ref{eq:ou}).
 Stars give the non-local contribution, appropriately normalized relative to 
 the open diamonds, i.e.,
 $\delta_{NL}\equiv (\Omega/4)\langle(r_1-r_2)\delta(v_{\phi,1}-v_{\phi,2})\rangle_c/(s_{\rm grid}\Omega)^2 $ (Eqs. [\ref{eq:f1}]--[\ref{eq:f2}]).  Bottom panel: post-collision relative velocities
 are indeed nearly isotropic, as we had surmised in Equation (\ref{eq:uu}). For this figure, we take
 $\bld{u}=\bld{V}_1-\bld{V}_2$, the total relative velocity; it includes the difference in the circular Keplerian
 velocities at the locations of the two particles.
\label{fig:micro}}
}
%%%%%%%%%%%%%%%%%%%%%%%%%%%%%%%%%%%%%%%%

\section{Summary and Outlook}
We have introduced an algorithm to simulate collisions between inelastic
particles in an optically thin disk orbiting a central mass.
The algorithm is simple to implement and adds negligible running time
to existing collisionless N-body codes. A major feature of the algorithm
is that the disk particles' motions need only be tracked in a plane.
Yet the algorithm transcends its two-dimensional appearance to simulate
a three-dimensional disk of particles whose random velocity distribution
tends to be isotropized by collisions.

We have performed a battery of tests of the algorithm for the case of an
isolated, narrow, circular ring. Numerical simulations agree with analytic
theory with regard to how the particles' velocity dispersion equilibrates,
how the ring viscously spreads, how energy and angular momentum are
transported, and how energy dissipation relates to the viscous
angular momentum flux and to the background shear. Angular momentum
transport arises not only from particle advection ($HF_n$), but also
from correlations in the random velocity field ($F_{r\phi}$) and from
finite particle sizes ($F_{\rm NL}$). The relative magnitudes of each of
these three terms can be measured from simulations. In making these
and other measurements, we sought ways to minimize noise introduced
by finite particle numbers (Poisson fluctuations). For example, when measuring
viscous fluxes of angular momentum and energy,
it proves useful to consider only
those particles that actually collide during the measurement interval.

The stage is now set for simulating more complicated systems---narrow
eccentric rings (like the Maxwell and Titan ringlets of Saturn, or the
Epsilon ring of Uranus), and circumstellar disks with embedded planets.
Among the phenomena we are interested in exploring numerically are the
formation of sharp edges by shepherd satellites, the evolution of narrow
rings into states of rigid apsidal precession, and the eccentricity evolution
of planets as driven by disks. 

\acknowledgements
We thank Ruth Murray-Clay for helpful exploratory calculations, and Jack Wisdom for
encouraging remarks.
EC acknowledges support from the National Science Foundation, NASA,
and the Alfred P.~Sloan Foundation, and is grateful for the warm hospitality
of the Canadian Institute for Theoretical Astrophysics /
University of Toronto, where a portion of this work was completed.

\vfill
\eject

\bibliographystyle{apj}
\bibliography{ms}

\end{document}